\documentclass[aps,prb,unsortedaddress,superscriptaddress,twocolumn]{revtex4-1}

\usepackage{xcolor}
\usepackage{epsfig}
\usepackage{graphicx}
\usepackage{amsfonts}
\usepackage[figuresright]{rotating}
\usepackage{amssymb}
\usepackage{amsmath}
\usepackage{color}
\usepackage{soul}
\usepackage{epstopdf}

\usepackage{graphicx}
\DeclareGraphicsExtensions{.eps, .PNG, .jpg, .pdf}

\newcommand{\ket}[1]{| #1 \rangle}
\newcommand{\bra}[1]{\langle #1 |}

\newcommand{\up}{\uparrow}
\newcommand{\down}{\downarrow}

\begin{document}

\title{Many-body mobility edge due to symmetry-constrained dynamics and strong interactions} 

\author{Ian Mondragon-Shem}
\affiliation{Department of Physics and Institute for Condensed Matter Theory, University of Illinois at Urbana-Champaign, Urbana, Illinois 61801}

\author{Arijeet Pal}
\affiliation{Department of Physics, Harvard University, Cambridge, Massachusetts 02138}

\author{Taylor L. Hughes}
\affiliation{Department of Physics and Institute for Condensed Matter Theory, University of Illinois at Urbana-Champaign, Urbana, Illinois 61801}

\author{Chris R. Laumann}
\affiliation{Department of Physics, University of Washington, Seattle, Washington 98195}

\date{\today}

\begin{abstract}
We provide numerical evidence combined with an analytical understanding of the many-body mobility edge for the strongly anisotropic spin-1/2 XXZ model in a random magnetic field.
The system dynamics can be understood in terms of symmetry-constrained excitations about parent states with ferromagnetic and anti-ferromagnetic short range order. 
These two regimes yield vastly different dynamics producing an observable, tunable many-body mobility edge.
We compute a set of diagnostic quantities that verify the presence of the mobility edge and discuss how weakly correlated disorder can tune the mobility edge further.
\end{abstract}

\maketitle
 

\section{Introduction} 
\label{sec:intro}

Many-body Anderson localization in isolated, disordered, interacting systems has been a topic of intense research in recent years. 
The localized phase is predicted to be a novel phase of matter which is characterized by its failure to thermalize under unitary quantum dynamics \cite{Anderson1958}. 
This breakdown of ergodicity endows the phase with a host of intriguing properties such as the absence of transport \cite{Basko2006, Gornyi2005}, 
protection of topological and Landau symmetry-breaking order \cite{Huse2013, Vosk2014, Chandran2014, Bahri2013}, and slow growth of entanglement \cite{Bardarson2012, Vosk2013, Serbyn2013a} even at \emph{infinite} temperature.
The initial discussions on this subject have been based on perturbation theory \cite{Fleishman1980, Basko2006, Gornyi2005}, 
numerical exact diagonalization \cite{Oganesyan2007, Pal2010, Iyer2013}, and real-space RG techniques \cite{Vosk2013,Vosk2014b,pekker2014}. 
Indeed, there is now a rigorous mathematical proof in one dimension \cite{Imbrie2014} and phenomenological descriptions \cite{Huse:2013uc,Serbyn2013b,Chandran:2014aa} of fully many-body localized systems at infinite temperature in terms of extensive collections of locally conserved quantities. 
However, much less is known regarding the properties of systems with a \emph{many-body mobility edge} at finite temperature, separating the localized and delocalized phases.

One of the outstanding issues for observing many-body localization (MBL) in experiments has been to differentiate it from single-particle Anderson localization (although see Refs. \onlinecite{Serbyn2014, Vasseur2014} for protocols in optical systems with local addressability).  
One approach is to detect a mobility edge as these do not exist for non-interacting systems in one dimension where all single particle states localize in the presence of disorder. 
Thus, in a non-interacting, localized system, all initial conditions lead to fully localized dynamics. 
In the presence of a mobility edge in an interacting system, on the other hand, correlations in the initial states, such as its energy, determine whether the system dynamics remain localized, or eventually thermalize.
Such initial condition dependent behavior provides an unambiguous signature of the many-body nature of the localization. 

The search for many-body mobility edges has recently intensified. They were perturbatively predicted several years ago in various models including weakly interacting fermions \cite{Basko2006} and continuum bosons in one dimension \cite{Aleiner:2010ir}. However, only recently have they begun to be numerically identified in certain spin models \cite{Jonas2014, Laumann2014, Luitz2014}. 
In the limiting case of a long-range mean-field model, a perturbative analysis provides a clear prediction of the mobility edge in quantitative agreement with numerical simulations \cite{Laumann2014}. 
Another recent study provided a compelling case for the presence of a mobility edge at the isotropic point of the XXZ model \cite{Luitz2014} through impressively large numerics.

In this article we provide numerical evidence, combined with an analytical understanding, of the many-body mobility edge in the strongly anisotropic spin-$1/2$ XXZ model in a random magnetic field. 
In the strongly interacting regime, the system dynamics can be understood in terms of excitations about parent states with ferromagnetic and anti-ferromagnetic short-range order. 
The dynamics of the excitations are strongly constrained by symmetry and, in particular, the ferromagnetic regime is much more susceptible to localization. 
Thus, preparing states with the appropriate correlations, and, tuning the short-range correlations in the disorder potential both provide knobs with which to explore the localization phase diagram.
Such tunability may be crucial for disentangling many-body effects from single-particle localization in experiments on finite-size systems of cold atoms \cite{Kondov2013}.

Remarkably, we provide an analytic picture of the finite temperature quantum transition that is quite different in essence from the seminal work of Basko, Aleiner, and Altshuler \cite{Basko2006}, as it builds on strongly interacting parent Hamiltonians, rather than non-interacting localized particles. 
The most important consideration of our analysis is that while the density of states of our parent Hamiltonian has a statistical symmetry when flipping the sign of the interaction strength, the low-energy dynamics in both regimes has a vastly asymmetric behavior. 
This is in sharp contrast to the unimportance of the sign of the interactions in the weakly interacting regime, and is crucial for the observation and tunability of the many-body mobility edge in our work.

Our article is organized as follows. 
In Section II, we define the model that we study to probe this phenomenon,  provide an analytical description for an asymmetry in the dynamics of excitations on top of parent ferromagnetic and anti-ferromagnetic orderings, and argue for the appearance of a many-body mobility edge. 
Section III introduces the diagnostic numerical quantities that we employ to observe the MBLD transition in numerical exact diagonalization calculations, and presents the numerical results.  
Finally, in Section IV, we discuss the introduction of correlated disorder as a means to manipulate the mobility edge.

\section{Dynamics in the random-field XXZ model} 
\label{sec:model}

We consider the one-dimensional spin-$1/2$ XXZ model in a random magnetic field with periodic boundary conditions:
\begin{equation}
H=\sum_{i=1}^{L}\left(t \left[\hat{S}^{+}_i  \hat{S}^{-}_{i+1}+ \hat{S}^{-}_i  \hat{S}^{+}_{i+1}\right]+U \hat{S}^{z}_i  \hat{S}^{z}_{i+1}+W  w_{i} \hat{S}^{z}_i\right). \label{XXZ}
\end{equation}
Here, $L$ is the number of lattice sites, $\hat{S}^{x,y,z}_i$ are the spin-$1/2$ operators ($\hbar = 1$), $\hat{S}^{\pm}_i=\hat{S}^x_i \pm i \hat{S}^y_i$  are the raising and lowering operators, and the couplings $w_{i}$ represent a short-ranged disorder potential. 
The XXZ model conserves the total spin projection $\hat{S}^z=\sum_{i}\hat{S}^{z}_i$.  
As we discuss below, this conservation law plays a fundamental role in inducing a mobility edge in the system.  

We assume throughout that the disorder ensemble satisfies the statistical symmetry $w_i \to - w_i$. 
Indeed, any statistically translationally invariant disorder ensemble may be shifted to have this symmetry at the expense of the introduction of a uniform field $[w] \sum_i \hat{S}^z_i$ (where we denote disorder averaged quantities by $[\ldots]$).
As this field couples to a conserved quantity, it has no effect on the dynamics and we may drop it.
For the numerics in Sec.~\ref{sec:numerics}, we take the  $w_i$  to be uniformly distributed in the range $[-0.5,0.5]$. This choice of disorder clearly possesses such a statistical symmetry. 
However,  the symmetry $w_i \to - w_i$ holds for more general disorder models, e.g. Gaussian disorder for which the $w_i$ have moments  $ [w_i] = 0$  and  $[ w_i w_{i'} ] =\delta_{i,i'}$. 

We will proceed by first noting that in the limit $t \rightarrow 0$ there is a symmetry between the ferromagnetic and antiferromagnetic regimes of the system. 
In particular, domain wall excitations in each type of ordering map into each other. 
Then we will consider the quantum dynamics of such domain-wall excitations mediated by a non-vanishing, but small, tunneling which is controlled by $t$. We will see that such dynamics behaves differently in the ferromagnetic and antiferromagnetic regimes due to the conservation of $\hat{S}^z$, which as a result creates a spectral asymmetry in the XXZ model. Such a spectral asymmetry was noted briefly in [\onlinecite{DeLuca2013}].

\begin{figure}
\begin{center}
\vspace{1.5cm}
\includegraphics[trim=0.cm 0cm 0cm 3cm, scale=0.23]{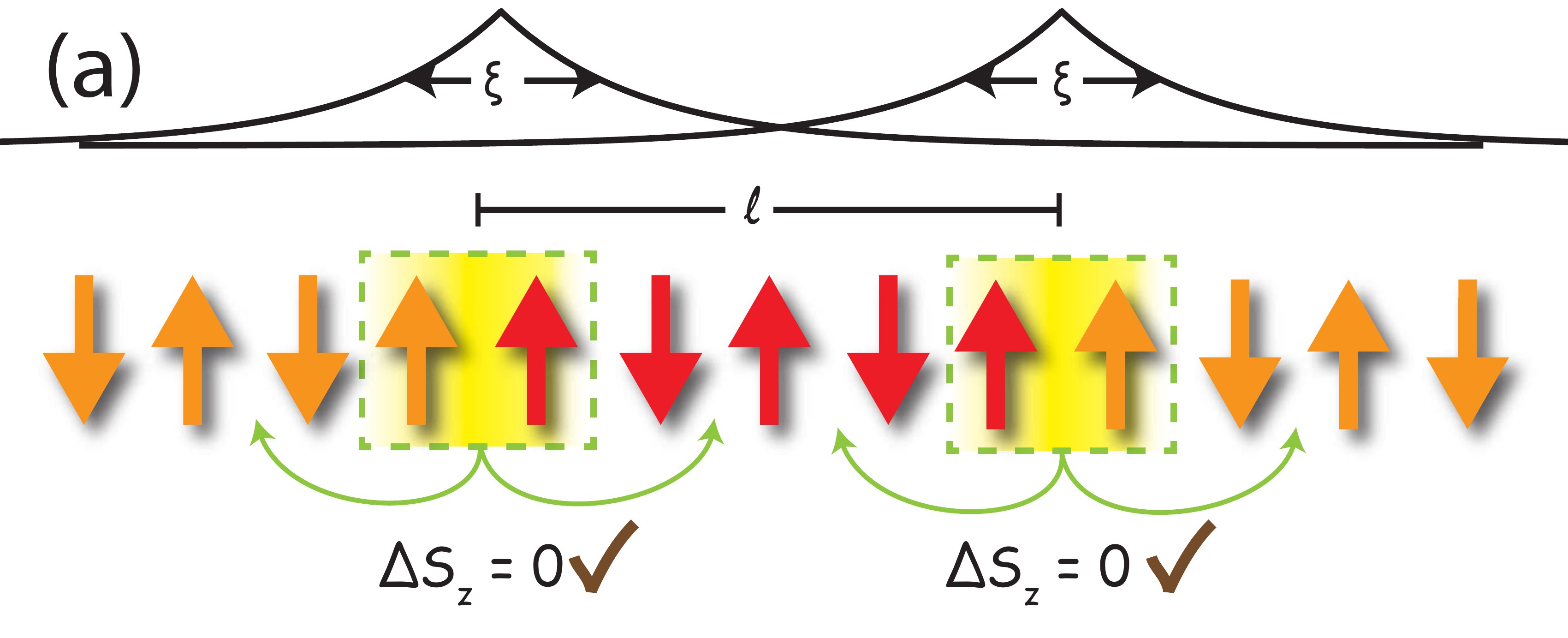}

\vspace{0.5cm}
\includegraphics[trim=0.cm 0cm 0cm 0cm, scale=0.23]{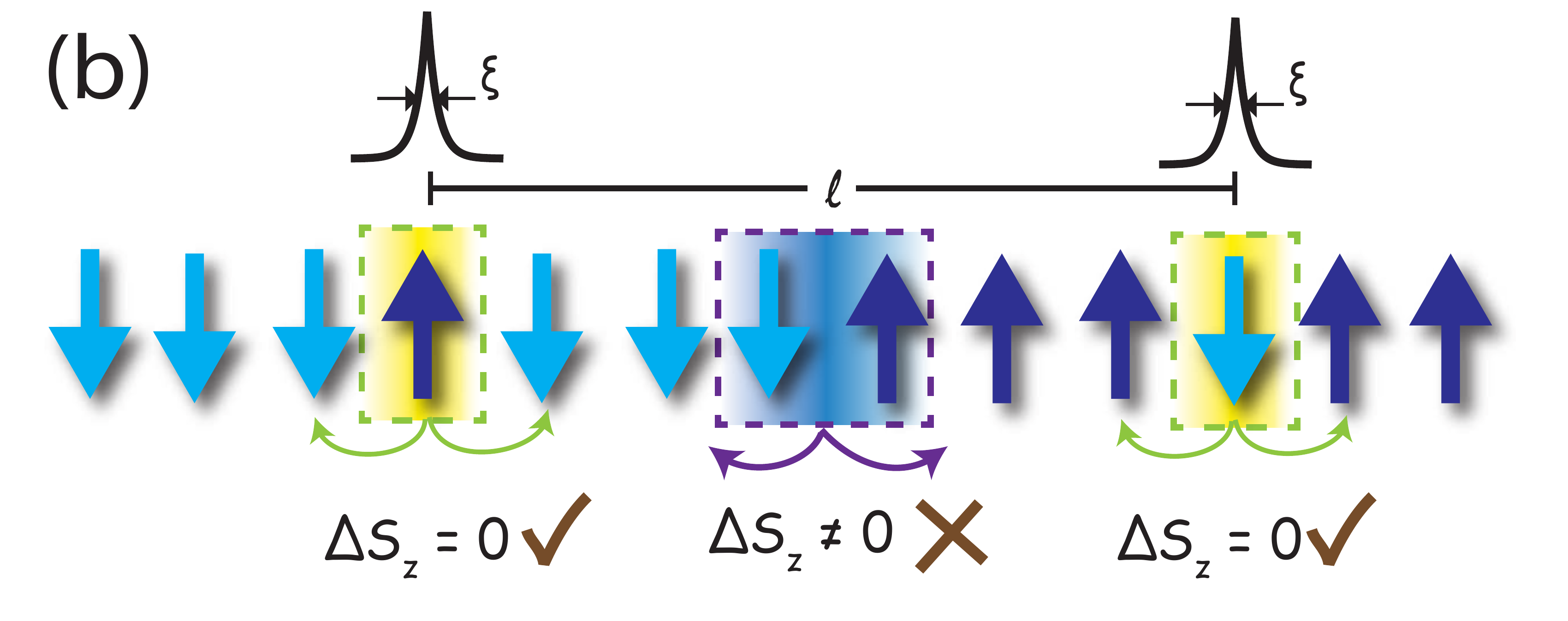}
\caption{(a) \textit{Dynamics in the antiferromagnetic regime:} Ne\'el domain walls are the lowest energy excitations (represented by green dashed squares) that can propagate because they conserve $S^z$. They move by two sites. (b) \textit{Dynamics in the ferromagnetic regime:} domain walls between ferromagnetic regions (represented by the purple dashed square) are not able to propagate due to the conservation of $S^z$. Higher energy excitations, namely single-spin flips (represented again by the green dashed squares), are the next available mobile excitations and can move by one site. These excitations  are more susceptible to localization due to the strong disorder potential.  In both regimes, the two important length scales associated with the mobile excitations are the localization length $\xi$ and the mean distance $\ell$ that separates them. A mobility edge forms when these two length scales become comparable.}\label{Fig_dyn}
\end{center}
\end{figure}

\subsection{Disorder-induced domain walls in the classical limit} 
\label{sec:strong_interactions}

In the limit $t\rightarrow 0$, the XXZ model becomes 
\begin{align}
        \label{eq:hamcl}
        H(U, W w_i) = U\sum_i S^z_i S^z_{i+1} + W \sum_i w_i S^z_i,
\end{align}
which is the well-studied classical random field Ising model in one dimension  \cite{Bruinsma:1983aa,Azbel:1983aa}. If we perform a Ne\'el transformation $\sigma^z_i \to (-1)^i \sigma^z_i$ on this model,
we obtain the mapping $H(U, W w_i) \to H(-U, (-1)^i W w_i)$. The transformed Hamiltonian $H(-U, (-1)^i W w_i)$  has the same statistical weight as $H(-U, w_i)$. Consequently, every state of the anti-ferromagnetic model $U>0$ can be mapped into a state of a ferromagnetic model with $U<0$ with the same energy and statistical weight. Thus, disorder averaged thermodynamic properties have $U\to -U$ symmetry, even though particular instances need not. It would be interesting to explore the effects of relaxing the constraint on the disorder distribution, such that this statistical symmetry was removed, but we leave these considerations to future work. 

Consider now the formation of domain walls due to the disorder potential in the ferromagnetic regime, i.e. the low (high) energy density regime when $U<0$ $(U>0)$.  A large interaction $\vert U\vert$ disfavors domain walls in the ground state. Naively, with $W=0$, the two symmetry breaking ground states of $H$ are simply the all $\up$ and all $\down$ configurations.
However, even infinitesimally small random fields destabilize these ferromagnetic ground states\cite{Nattermann1997}, as usual for $d \le 2$. 
The random field $W$ lifts the ground state degeneracy by an energy typically of order $E_\down - E_\up = W \sum_{i} w_i \sim \sqrt{L} W $ where $L$ is the length of the chain (in lattice units). 
More generally, any domain of length $l$ acquires a random field energy of order $\sqrt{l} W $, which, for $l$ large enough, always exceeds the energy cost $|U|/2$ per domain wall. 
Thus, the ground state of the random field system builds in a collection of domain walls separated by random lengths typically of the scale $l_0 \sim (U/W)^2$. 
The actual ground state has vanishing magnetization density (with $1/\sqrt{L}$ fluctuations), and consists of pinned alternating domains of length $l$.

Due to the statistical Ne\'el symmetry, we can immediately conclude that in the anti-ferromagnetic regime ($U>0$), the formation of domain walls is the same as  the ferromagnetic case. All one has to note is that the Ne\'el transformation maps the two types of ferromagnetic domains into the two types of possible antiferromagnetic domains, namely those for which the $\up$ spins live on either the even or odd sublattice. The domain walls in this case are then phase slips in this Ne\'el ordering. The ground state is thus composed of a dilute gas of Ne\'el domain walls due to the random field. The density of these domain walls is $n^0_{DW} = 1 / l_0 \sim (W/U)^2$.  We conclude then that in the limit $t\rightarrow 0$ there is a clear statistical symmetry between the ferromagnetic and antiferromagnetic regimes.

\subsection{Spectral asymmetry in the quantum dynamics of domain wall excitations} 
\label{sub:quantum_dynamics}

We now discuss the low-energy quantum dynamics of domain walls in the limit where $0<|t| \ll |W| \ll |U|$. 
By comparing the localization length of the available mobile excitations to their mean separation, we will obtain an estimate of the temperatures $T_{\pm}$ for which a mobility edge forms in the antiferromagnetic and ferromagnetic regimes, respectively. 
Despite the statistical symmetry we discussed in the previous section, it turns out that the ferromagnetic and antiferromagnetic regimes have different types of available mobile excitations essentially due to the conservation of $S^z$. 
This leads to different transition temperatures. 

\subsubsection{Antiferromagnetic regime} 
\label{ssub:antiferromagnetic_limit}

Consider the antiferromagnetic regime ($U>0$) at low energy densities. Let us represent a spin state with a single domain wall  between two Ne\'el domains as \begin{align}\ket{\cdots\up\down\up\down \vdots \down \up \down \up \cdots},\nonumber\end{align} where the vertical dots denote the domain wall itself.  The tunneling term of the Hamiltonian, i.e. the term proportional to $t$, allows for such Ne\'el domain walls to hop freely by two lattice sites:
        \begin{align}
              \ket{\cdots\up\down\up\down \vdots \down \up       \down \up \cdots} &\,\xrightarrow{\,\,t\,\,}\,
              \ket{\cdots\up\down\up\down        \up \down \vdots \down \up \cdots} \nonumber.
        \end{align}
Additionally, they can annihilate if they belong to the same sublattice:
        \begin{align}
                \ket{\cdots\up\down\up\down \vdots \down \up  \vdots   \up  \down \cdots} &\,\xrightarrow{\,\,t\,\,}\,
                \ket{\cdots\up\down\up\down        \up \down  \up \down \cdots}\nonumber.
        \end{align}
There are thus two flavors of domain walls that propagate throughout the system -- those which live on the even and odd sublattices.  In addition to this, $S^z$ conservation prevents the even and odd flavors from passing through each other as we can see from the following configurations belonging to different $S^z$ sectors:
        \begin{align}
                \begin{array}{lr}
                \ket{\cdots\up\down\up\down \vdots_e \down \vdots_o \down \up  \cdots} & \Rightarrow  S^z = -1, \nonumber\\
                \ket{\cdots\up\down\up\vdots_o \up \vdots_e \up \down \up  \cdots} & \Rightarrow  S^z = +3.\nonumber 
                \end{array}
        \end{align}
Note from these properties that the dynamical behavior of domain walls in the XXZ model is very different from, for example, that of the transverse field Ising model without the addition of a $U(1)$ symmetry such as the conservation of $S^z$. In the transverse field Ising model, the lack of $S^z$ conservation allows domain walls to move by a single lattice site and to pass through each other because there can be single spin flips at a time. 

Based on these general observations, we can think of the domain walls as particles that hop with strength $t$ by two lattice sites at a time, and which feel a disordered potential of strength $\sim W$. With this picture in mind, the effective localization length of a single domain wall can be estimated to be of order $\xi \approx 2/\ln(\sqrt{2}W/t)$.  On the other hand, to obtain the average inter-particle spacing we must take into account two possible contributions to the density of domain walls. As we stated in the previous section, the ground state already has a density of domain walls $n^0_{DW} $ built-in due to the disorder potential. In addition to this, as the temperature $T$ increases, the density of excess domain walls increases by an amount $n^{exc}_{DW} \sim e^{-|U|/2T}$. Hence, the average separation between domain walls is $\ell(T) = \left(n^0_{DW} + n^{exc}_{DW}\right)^{-1}$. We illustrate these length scales in Fig.~\ref{Fig_dyn}a.

The transition temperature that determines the mobility edge is given by the condition $\xi \approx  l(T)$. To understand this, note that when the domain walls come into contact, they interact with a very large energy $U.$  Thus, at arbitrarily small temperatures, delocalization should occur immediately if $\xi > \ell(T)$ such that there is more than one excitation per localization volume built into the ground state. Furthermore, such delocalization should persist as temperature increases. If, on the other hand,  $l(T) \gg \xi$,  the domain walls do not interact and the state is many-body localized. Hence, we expect the boundary between these two types of behavior to occur when $\xi \approx  l(T)$, which leads to:
\begin{align}
  \frac{2}{\ln \left(\sqrt{2} W / t\right)} \approx \frac{1}{(W/U)^2 + e^{-|U|/2T_+}}.\label{antiferro}
\end{align}
For fixed values of $W$ and $U,$ this expression provides an estimate of the temperature $T_+$ for which a mobility edge is expected to emerge on the antiferromagnetic side of the energy spectrum. This means that starting from the regime $U>0$ at low-temperature we would expect a finite-temperature transition at $T_{+}$ from a MBL phase to a delocalized phase.   

\subsubsection{Ferromagnetic regime} 
\label{ssub:antiferromagnetic_limit}

Let us now discuss the localization properties of the ferromagnetic regime ($U<0$) at low energy density. The main difference with the antiferromagnetic case is that, even though domain walls between different ferromagnetic regions naturally populate the ground state, the conservation of $S_z$  does not allow them to propagate throughout the system. Take, for example, the following configurations:
\begin{eqnarray}
\ket{\cdots \down \down \vdots \up \up \cdots} & \Rightarrow& S^z = 0, \nonumber\\
\ket{\cdots \down \vdots \up \up \up \cdots} & \Rightarrow& S^z = +2, \nonumber\\
\ket{\cdots \down \down \down \vdots \up  \cdots} & \Rightarrow& S^z = -2. \nonumber 
\end{eqnarray}
It is clear that for the domain wall in these examples to move one would have to violate the conservation of $S^z$. The least massive mobile excitations in this regime are actually domains of length $1$, i.e. magnons or spin flips. These excitations hop with strength $t$ by one lattice site, but they cost $U$ interaction energy to produce or destroy. Additionally, any anomalously short domain built into the ground state is necessarily in an anomalously strong field and is unable to move.  

The localization length for a mobile magnon in the $W \gg t$ limit is of order $\xi \approx \frac{1}{\ln\left(W/t\right)}$. Furthermore, their mean separation is set entirely by their thermal population $\ell(T)=n^{-1}_{M} \approx e^{|U|/T}$. We illustrate these length scales in Fig.~\ref{Fig_dyn}b. It thus follows that the mobility edge in the ferromagnetic regime is given by the condition $\xi \sim \ell(T_-)$, which yields
\begin{equation}
\frac{1}{\ln\left(W/t\right)} \sim e^{|U|/T_-}. \label{ferro}
\end{equation}
This expression determines the temperature $T_-$ at which a mobility edge forms in the ferromagnetic regime. 

\subsubsection{Appearance of a mobility edge} 
\label{ssub:mobility edge}

The different dynamical behaviors exhibited in the ferromagnetic and antiferromagnetic regimes can sharpen the signatures of a mobility edge in the XXZ model. To be concrete, for all of our calculations we will fix $U>0.$ This means that, without disorder, the ground state of the model is antiferromagnetic, whereas the highest energy state is ferromagnetic. It follows that the states near the bottom part of the energy spectrum are excitations on top of an antiferromagnetic parent ground state, and the states near the highest part of the energy spectrum are effectively excitations on top of a ferromagnetic  state. Then, depending on whether we are at positive or negative temperature determines which dynamical regime dominates the physics. For finite positive (negative)  temperatures we will be in the antiferromagnetic (ferromagnetic) regime. Thus, the dynamical asymmetry is present in a single model with a fixed $U$, and realized by the fact that different sections of the energy spectrum have quite different dynamics. 

Now consider increasing the disorder strength starting from zero. The states near both ends of the energy spectrum generically localize first, i.e. before the states in the middle of the band (near energy densities corresponding to infinite temperature). Because of this, two mobility edges generically form at temperatures $T_{\pm}$ below and above infinite temperature. This was observed numerically, for example, at the isotropic point $t=2U$ that was studied in [\onlinecite{Luitz2014}], where it was found that $T_+\sim T_-$. In finite-size numerical studies, the fact that two mobility edges form simultaneously makes it difficult to observe a sharp mobility edge because the thermal part of the spectrum is reduced from both sides of the spectrum. In practice, it is observed that either the thermal or MBL parts of the spectrum do not clearly stabilize.

In the anisotropic case, however, the two mobility edges occur at different temperatures. Comparing Eq.~\ref{antiferro} with Eq.~\ref{ferro}, one can see that, if $W$ and $U$ are chosen appropriately, the additional density of mobile excitations in Eq.~\ref{antiferro} can lead to $T_{-}>T_+$. In other words, we can have a situation where the disorder is strong enough that it localizes the ferromagnetic part of the spectrum, but weak enough that the antiferromagnetic side remains thermal. A clear mobility edge should thus arise. Let us now show numerically that such an asymmetry does make it easier to observe the mobility edge in this strongly interacting system.

\section{Numerical evidence} 
\label{sec:numerics}

After an extensive numerical exploration of the parameter space determined by $(U/t, W/t, S_z)$, we have found that the mobility edge is most strongly stabilized when interactions and disorder are strong with respect to $t$, and there is vanishing spin-polarization in the system $S^z=0$. This limit is compatible with our qualitative analysis in the previous section, and we can apply the intuition developed there to understand the numerical results. Interestingly, this $S^z$ sector is equivalent to a half-filled model of interacting fermions via the Jordan-Wigner transformation. We thus expect our conclusions to apply to such a fermionic system as well. 

To numerically ascertain the presence of the mobility edge, we performed exact diagonalization of the Hamiltonian defined in Eq.~\ref{XXZ}. Using the many-body energy densities, and corresponding energy eigenstates of the system, we computed three diagnostic quantities that have been used in the past to determine whether a system is in a thermal or in a many-body localized phase, although here we use energy-density resolved versions of these diagnostic tools. 

In order to obtain stable statistical behavior in these diagnostic quantities for a given energy density $E/L$, it is convenient to perform spectral and disorder averages within a set of states with energies close to $E/L$. To do this, first we shifted and rescaled the energy\cite{Jonas2014} as $\epsilon=\left(E-E_{0}\right)/\Omega$, where $\Omega$ is the energy bandwidth and $E_0$ is the ground state energy. This means that, by definition, $\epsilon$ ranges from $0$ to $1$. However, since the density of states is very low at the band edges, we sampled energies only in the range $\epsilon \in [0.1,0.9]$. For a given value of $\epsilon$, we obtained the set  $\{\epsilon_n\}_{\epsilon}$ of $50$ closest many-body energy eigenvalues. We then computed each of the four diagnostic quantities within the set $\{\epsilon_n\}_{\epsilon}$, and subsequently carried out spectral and disorder averages to obtain the final result. In what follows, we will denote spectral and disorder averages as $[ \langle \ldots \rangle ]$. 

\begin{figure}
\begin{center}
\includegraphics[trim=0.5cm 0cm 0cm 0cm, scale=0.35]{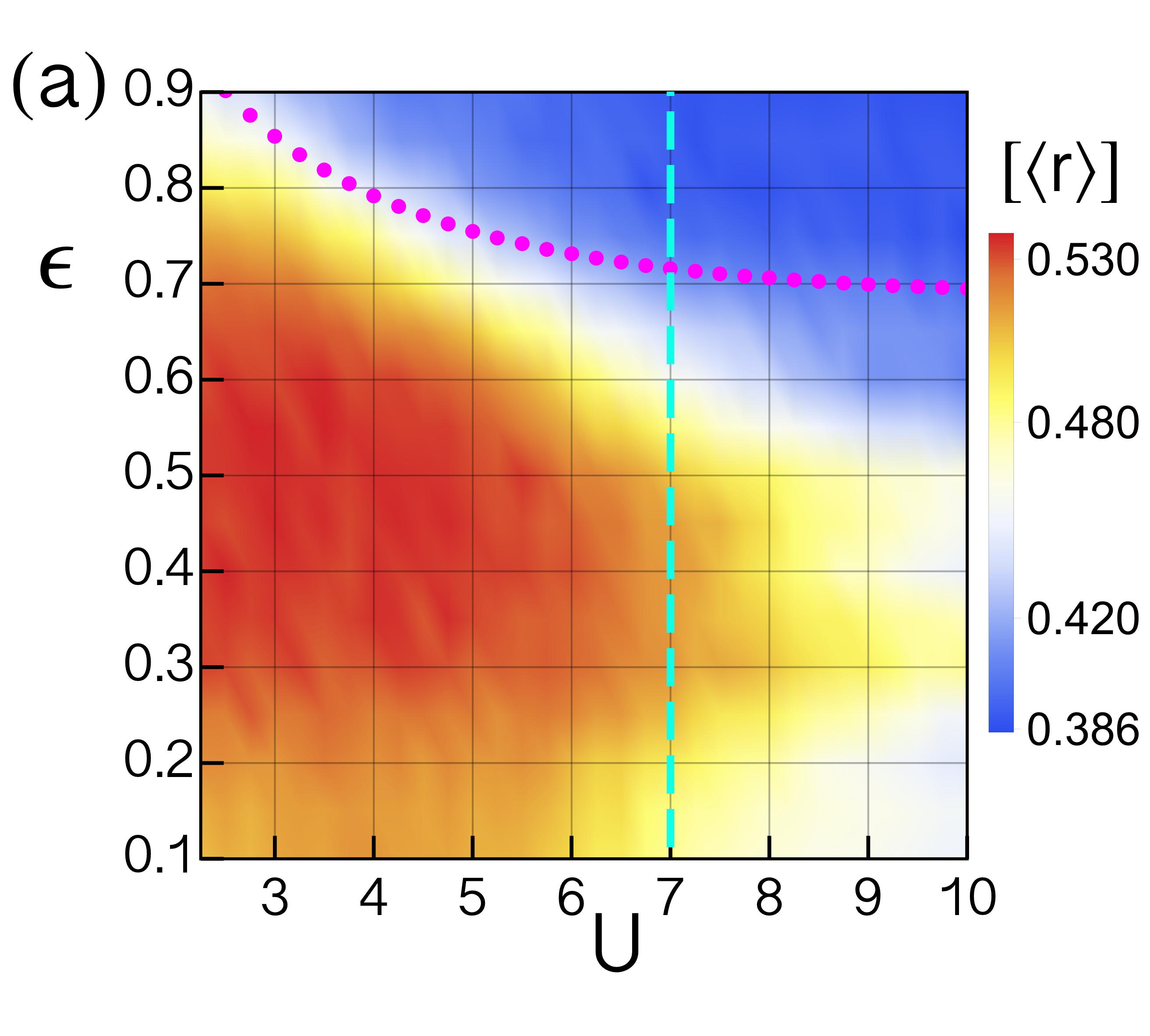}
\includegraphics[trim=0.5cm 0cm 0cm 0cm, scale=0.35]{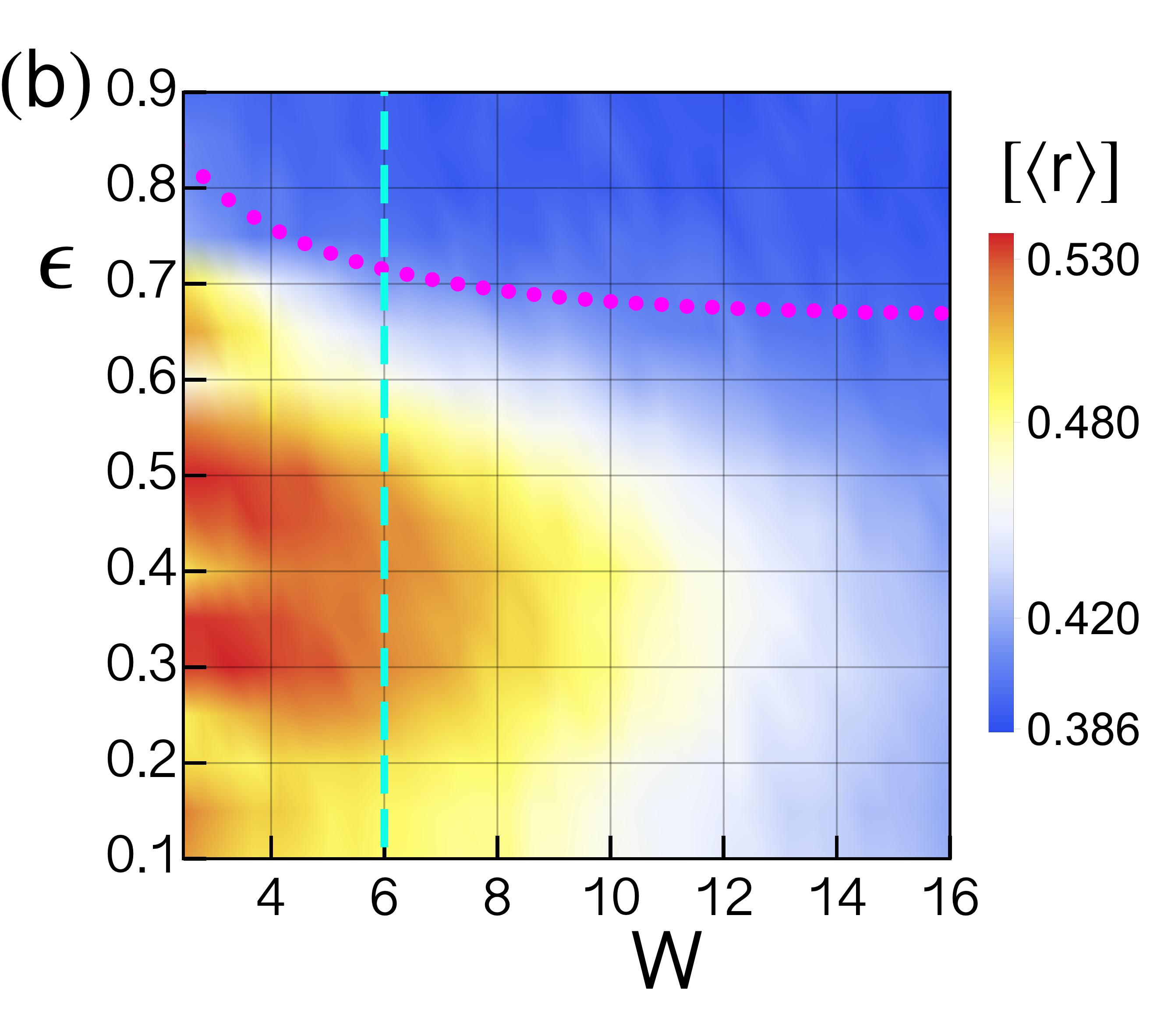}
\caption{Intensity plots of $[\langle r \rangle]$ as a function of:  (a) interaction strength $U$ with fixed $W=6.0$, and (b) disorder strength $W$ with fixed $U=7.0$. The magenta points show the fit of the mobility edge $\epsilon_c(U,W)$ using Eq.~\ref{edge}. The vertical blue lines denote the particular case for which the other diagnostic calculations are performed and shown in Fig.~\ref{Fig_scal_uncorr}. We used $500$ disorder realizations and $L=14$ for both cases. We caution that the mobility edge is not necessarily the white colored region, but rather we estimate it to be within the region for which $[\langle r \rangle]\in (0.42,0.48)$.}\label{Fig_PD_uncorr}
\end{center}
\end{figure}

We start with the energy-resolved statistics of many-body energy level spacings. We calculated the ratios
\begin{equation}
r_n=\frac{\text{min}(\delta_n,\delta_{n-1})}{\text{max}(\delta_n,\delta_{n-1})},
\end{equation}
where $\delta_n = \epsilon_{n}-\epsilon_{n-1}$, and where $\{\epsilon_n\}_{\epsilon}$ is assumed to be sorted in ascending order. In the MBL phase, the energy level spacings satisfy $[\langle r \rangle]_{MBL}\approx 0.386$, which is the average of a Poisson probability distribution. In the thermal phase, since the model we are considering is in the GOE, one should obtain $[\langle r \rangle]_{T}\approx 0.529$. Using these limits we can determine whether the eigenstates for a given range of energy densities are thermal or many-body localized. 

In Fig.~\ref{Fig_PD_uncorr}a we show an intensity plot of $ [\langle r \rangle ]$ as a function of interaction strength for a fixed value of disorder strength $W=6.0$.  Red regions denote values of $ [\langle r \rangle ]$ that are close to $[\langle r \rangle]_{T}$, whereas blue regions denote values close to $[\langle r \rangle]_{MBL}$.  Similarly, in Fig.~\ref{Fig_PD_uncorr}b we show $ [\langle r \rangle ]$ as a function of disorder strength with fixed interaction strength $U=7.0$. Both figures were computed with $L=14$ and $500$ disorder realizations. The boundary between $[\langle r \rangle]_{T}$ and $[\langle r \rangle]_{MBL}$ clearly reveals the existence of a mobility edge $\epsilon_c(U,W)$ that is dependent on both disorder and interaction strengths. More importantly, this figure shows that the ferromagnetic states (i.e. those at high energy density) are more susceptible to localization than the antiferromagnetic states (i.e. those at low energy density), consistent with our previous discussion. We caution, however, that the mobility edge is not necessarily the white colored region, but rather we roughly estimate it to be within the region for which $[\langle r \rangle]\in (0.42,0.48)$.

As a check that our physical picture is reasonable, we can obtain an estimate of $\epsilon_c(U,W)$ using  Eq.~\ref{ferro}. Let us write this condition in the form
\begin{equation}
\frac{1}{\ln \left(W/t\right)}=\gamma_1 e^{U/ T }\label{cond},
\end{equation}
where we take $\gamma_1$ is a fitting parameter that tunes the density of mobile excitations. In order to write the temperature in terms of energy density, we make use of the relation
\begin{equation}
\frac{1}{T}=\frac{1}{\Omega(U,W)}\frac{\partial S}{\partial \epsilon},
\end{equation}
where $S\propto \ln \rho(\epsilon)$ is the entropy and $\rho(\epsilon)$ is the disorder-averaged density of states. Note that we have made explicit the fact that the energy band-width is in general dependent on both $U$ and $W$. We now approximate the density of states by a Gaussian distribution
\begin{equation}
\rho(\epsilon)=\mathcal{N} e^{-\alpha(U,W)\left(\epsilon-\epsilon_0(U,W)\right)^2}.
\end{equation}
where $\mathcal{N}$ is a normalization factor, $\alpha^{-1}(U,W)$ controls the width of the distribution and $\epsilon_0(U,W)$ is the energy density at infinite temperature. These two numbers depend in general on both $U$ and $W$. The explicit relation between $T$ and $\epsilon$ is then given by
\begin{equation}
\frac{1}{T}=-\frac{2\alpha(U,W)}{\Omega(U,W)}\left(\epsilon_0(U,W)-\epsilon\right), \label{Te}
\end{equation}
We have introduced a minus sign in this expression to take into account that the ferromagnetic regime in the numerics actually occurs when $\epsilon_0(U,W)<\epsilon$. By replacing Eq.\ref{Te} in Eq.\ref{cond} and solving for $\epsilon$, we finally obtain
\begin{equation}
\epsilon_c(U,W)=\epsilon_0(U,W)-\frac{\Omega(U,W)}{\alpha(U,W)}\left[\frac{\ln \ln W+\ln \gamma_1}{2U}\right].\label{edge}
\end{equation}
Note that both $\epsilon_0(U,W)$ and $\alpha(U,W)$ can be calculated by numerically fitting $\rho(\epsilon)$ for each $U$ and $W$ that is sampled.  The bandwidth $\Omega(U,W)$ can also be obtained from the calculated energy spectrum. There is then only one free parameter that we can adjust, namely $\gamma_1$. 

In particular, using $\gamma_1= 0.2$ and $N=14$, we obtain the magenta points in both Figs.~\ref{Fig_PD_uncorr}a,b. The overall trend of this estimate of $\epsilon_c(U,W)$ qualitatively tracks the numerically obtained mobility edge. It is interesting to note that both the dependence on disorder ($W$) and interaction strength ($U$) of the many-body mobility edge are in reasonable quantitative agreement with only a limited choice of free parameters.  

\begin{figure}
\begin{center}
\includegraphics[trim=0.cm 0.5cm 0.5cm 0cm, scale=0.358]{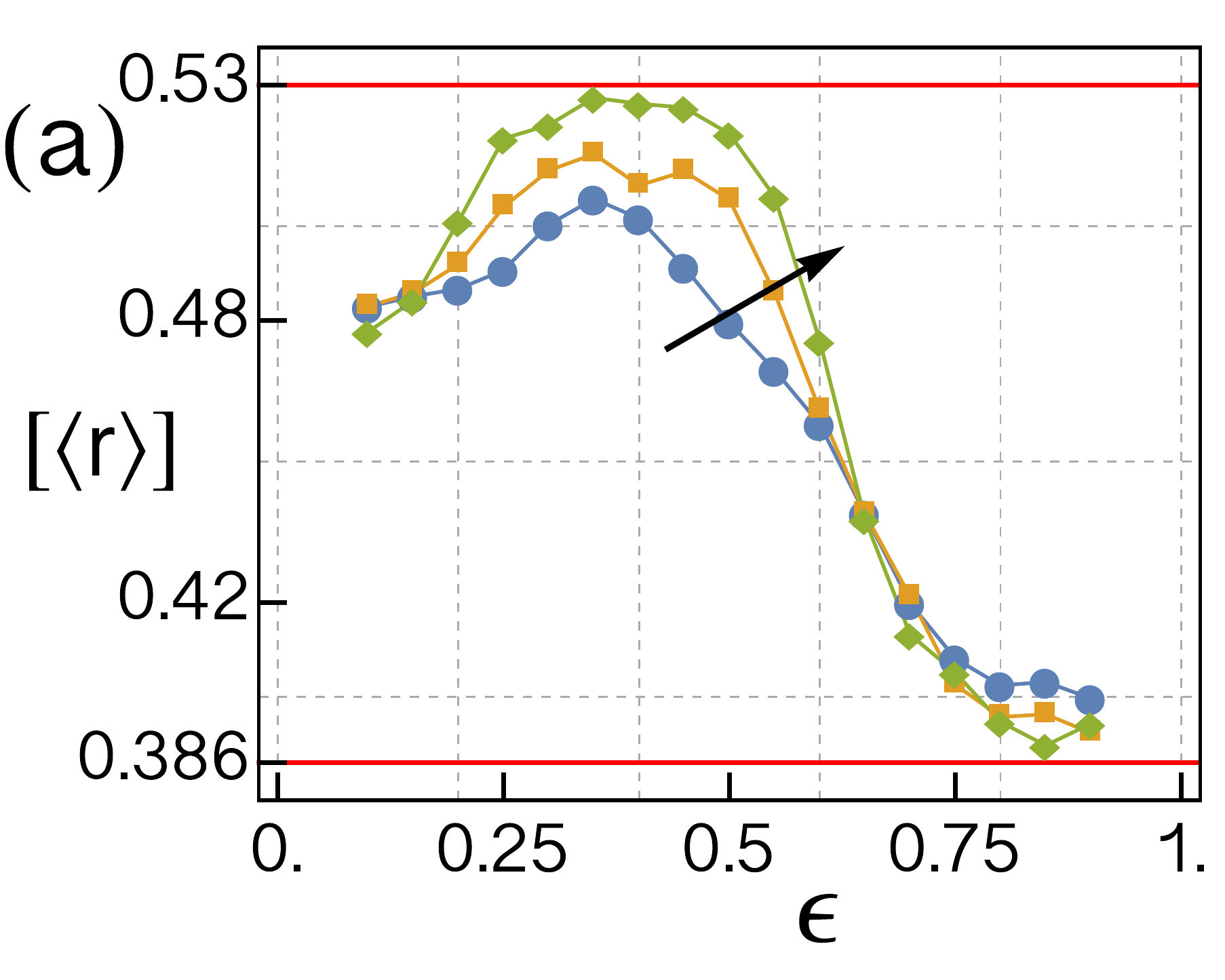}
\includegraphics[trim=0.cm 0.5cm 1.5cm 0cm, scale=0.355]{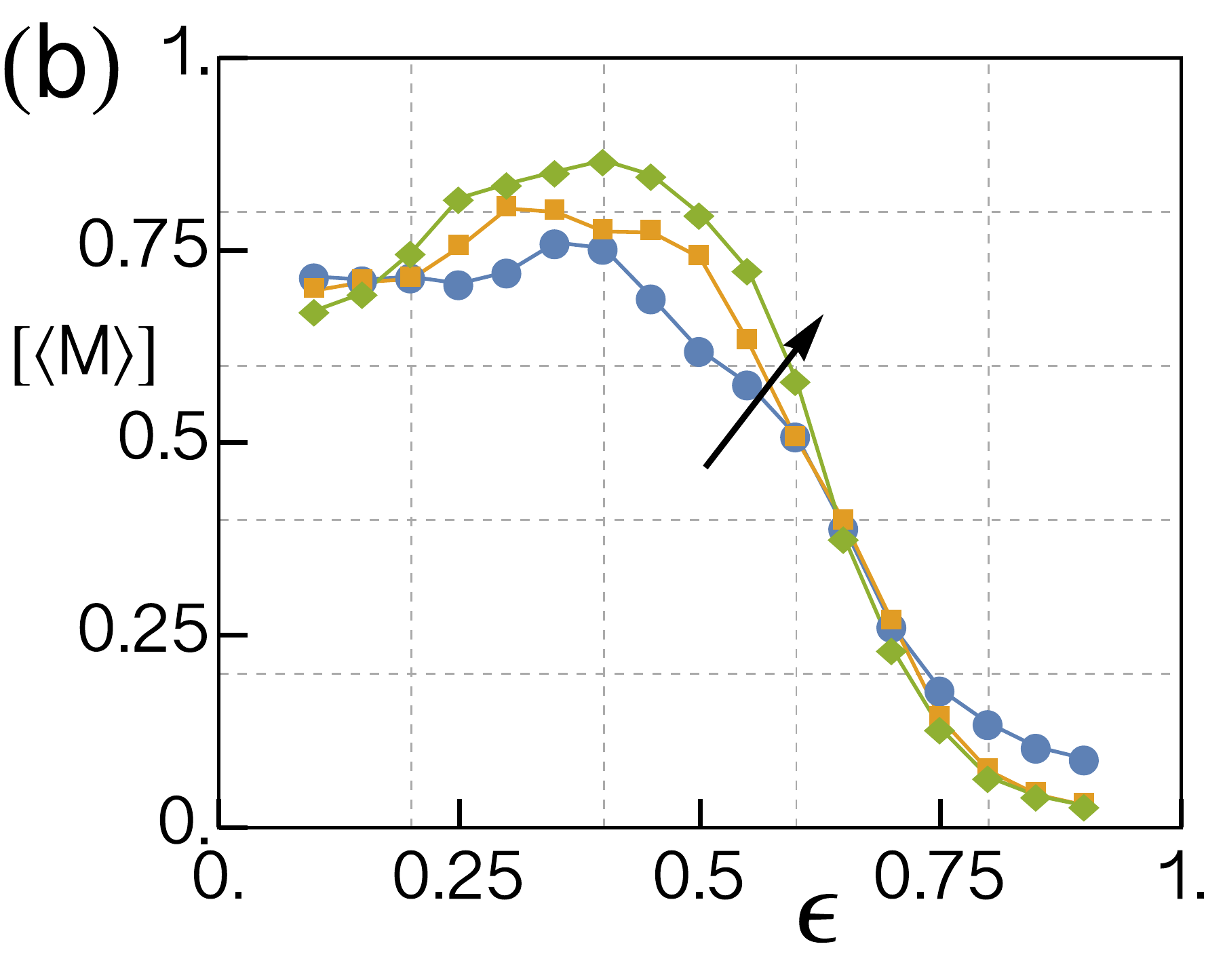}
\includegraphics[trim=0cm 1.5cm 0.cm 0cm, scale=0.355]{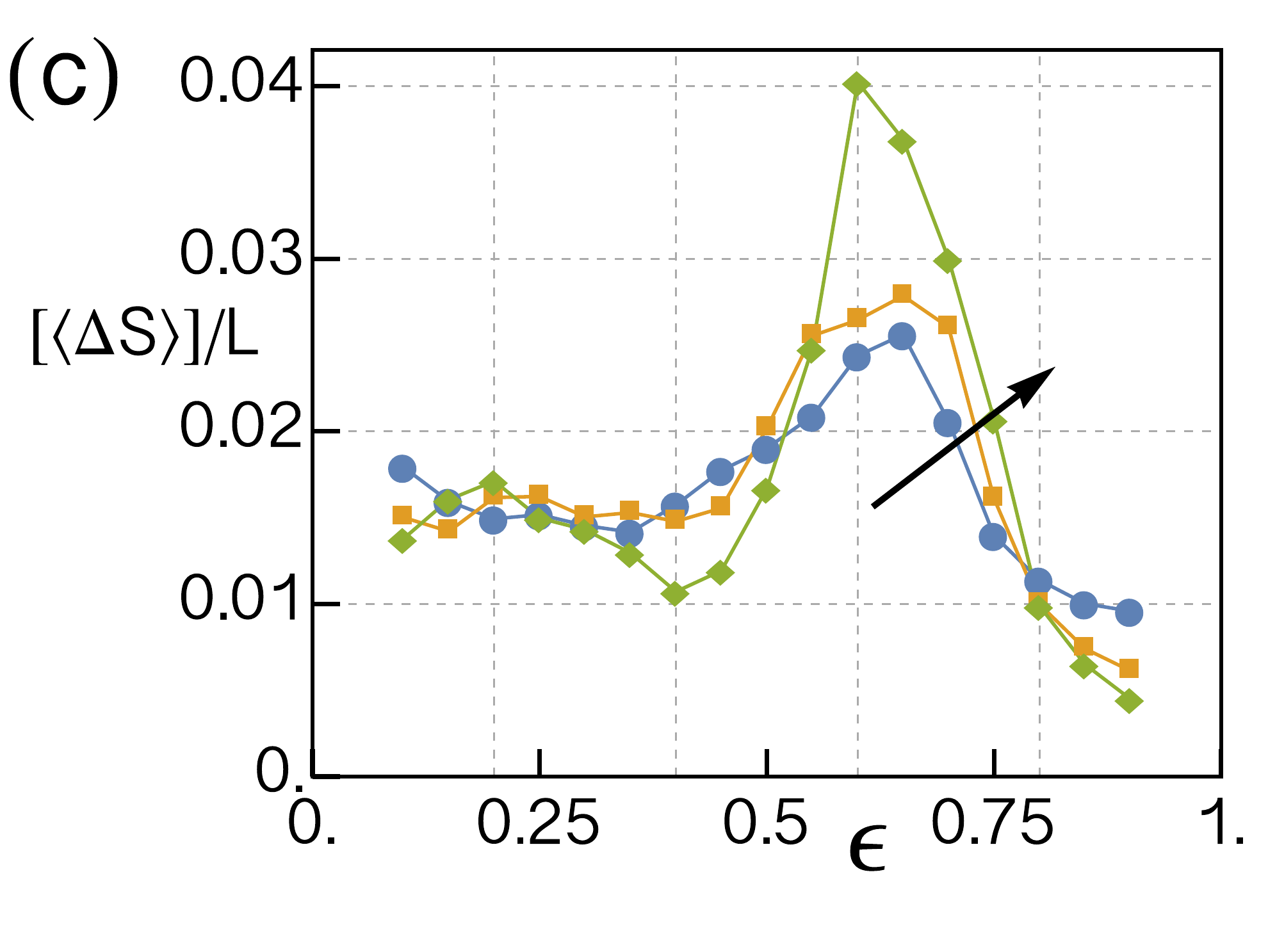}
\caption{Numerical evidence of the mobility edge in the spin-$1/2$ XXZ model: (a) Statistics of many-body energy level spacings $[\langle r\rangle]$. (b) Fraction of initial spin density modulation which is dynamic $[\langle M\rangle]$. (c) Variance of the entanglement entropy $[\langle \Delta S\rangle]$. Each of the curves in the figures corresponds to the system sizes $L=12\,(\text{blue}),14\,(\text{yellow}),16\,(\text{green})$.  The arrows denote the increase in system size. The parameters used are $(U,W,S_z)=(7.0,6.0,0)$. The number of disorder realizations used was 500 for all the curves, except for the entanglement entropy of the $L=16$ system, for which the number of disorder realizations used was $150$.}\label{Fig_scal_uncorr}
\end{center}
\end{figure}

Let us now focus on the particular case $(U,W,S^z)=(7.0,6.0,0)$, which corresponds to the vertical blue lines in Figs.~\ref{Fig_PD_uncorr}a,b. In Fig.~\ref{Fig_scal_uncorr}a, we show $ [\langle r \rangle ]$ for three system sizes, $L=12, 14, 16$. As the normalized energy density is traversed from $0$ to $1$, the averaged ratios $[\langle r \rangle]$ transition from $[\langle r \rangle]_{T}$ to $[\langle r \rangle]_{MBL}$ in the neighborhood of $\epsilon\approx 0.7.$ Importantly, as the system size is increased, the transition becomes sharper, as would be expected for this to be a true mobility edge in the thermodynamic limit. Finally, note that near the very bottom of the energy band, the $[\langle r \rangle]$ value decreases again, which could indicate a mobility edge forming on the antiferromagnetic side, but at a much lower temperature.

We can also probe the many-body eigenstates in order to characterize this transition further.  We start by considering the spin transport properties of the eigenstates by computing the relaxation behavior of an initial modulation of the spin density. We introduce such a modulation with the operator
\begin{equation}
\hat{M}=\sum_{j}\hat{S}^z_j e^{i 2\pi j/L }.
\end{equation}
At long times, the fraction that remains dynamic, given a many-body eigenstate $\ket{\Omega_m}$, is \cite{Pal2010}
\begin{equation}
M_m= 1-\frac{\bra{\Omega_{m}} \hat{M}^{\dagger}\ket{\Omega_{m}}\bra{\Omega_{m}} \hat{M}\ket{\Omega_{m}}}{\bra{\Omega_{m}} \hat{M^{\dagger}}\hat{M}\ket{\Omega_{m}}}.
\end{equation}
It is expected that in the MBL phase, where relaxation is not possible, $M_m$ should be vanishingly small. By contrast, in the thermal phase $M_m$ should approach unity, meaning full relaxation of the modulated state. In Fig.~\ref{Fig_scal_uncorr}b, we show the behavior of $[\langle M\rangle]$ for the present model. This figure shows that the modulation is able to relax for states within the energy range $0.25<\epsilon<0.6,$ whereas for $\epsilon>0.6$ this relaxation is suppressed. Furthermore, this behavior clearly has the appropriate finite-size scaling, and from the finite-size scaling trend the deviation from the expected infinite-size limit of $1$ is ascribed to be a finite-size effect.

Finally, we computed the entanglement properties of the system. The typical measure of entanglement is the von-Neumann entropy for a given eigenstate $\ket{\Omega_n},$ and is given by
\begin{equation}
S^{(n)}_A=-\text{Tr}_L\left(\rho^{(n)}_L \log \rho^{(n)}_L\right),
\end{equation}
where $\text{Tr}_L$ represents the trace over the degrees of freedom in the left region and $\rho^{(n)}_L=\text{Tr}_{R}\left(\ket{\Omega_n}\bra{\Omega_n}\right)$ is the reduced density matrix of the left region. Instead of the entropy itself, we calculated the disorder-induced variance $\Delta S$ of the entanglement entropy. This quantity was argued to show a peak at the many-body mobility edge. The reason being that, for a finite-size system,  a small window of energy density that contains the critical energy density will contain thermal and MBL states. Thus, the entanglement entropy fluctuates between volume and area law scaling, as a result producing a large variance in the distribution of entanglement entropy \cite{Jonas2014}.  In Fig. \ref{Fig_scal_uncorr}c, we show $\Delta S/L$ as a function of energy density. A peak emerges near $\epsilon\sim 0.6$, which becomes clearer as the system size increases, lending further support for a true mobility edge forming in the neighborhood of these energies. 

To summarize, the numerical results presented here  all confirm that the ferromagnetic regime shows a much greater affinity for many-body localization than the anti-ferromagnetic regime, which is consistent with our analysis of their asymmetric dynamics. It appears that the anti-ferromagnet can delocalize at low-temperatures while the ferromagnet requires a much higher (negative) temperature to undergo a transition into the thermal phase. This leads to the presence of a MBLD transition which shifts asymmetrically away from the center of the band, i.e., away from infinite temperature, as a function of the relative disorder strength, which is a distinct feature of the many-body nature of the phenomena. With these promising results we will now introduce correlated disorder with the intention of creating even stronger asymmetry in the dynamics, and thus strengthening the signature of the many-body mobility edge. 

\begin{figure}
\begin{center}
\includegraphics[trim=0.5cm 0cm 0cm 0cm, scale=0.35]{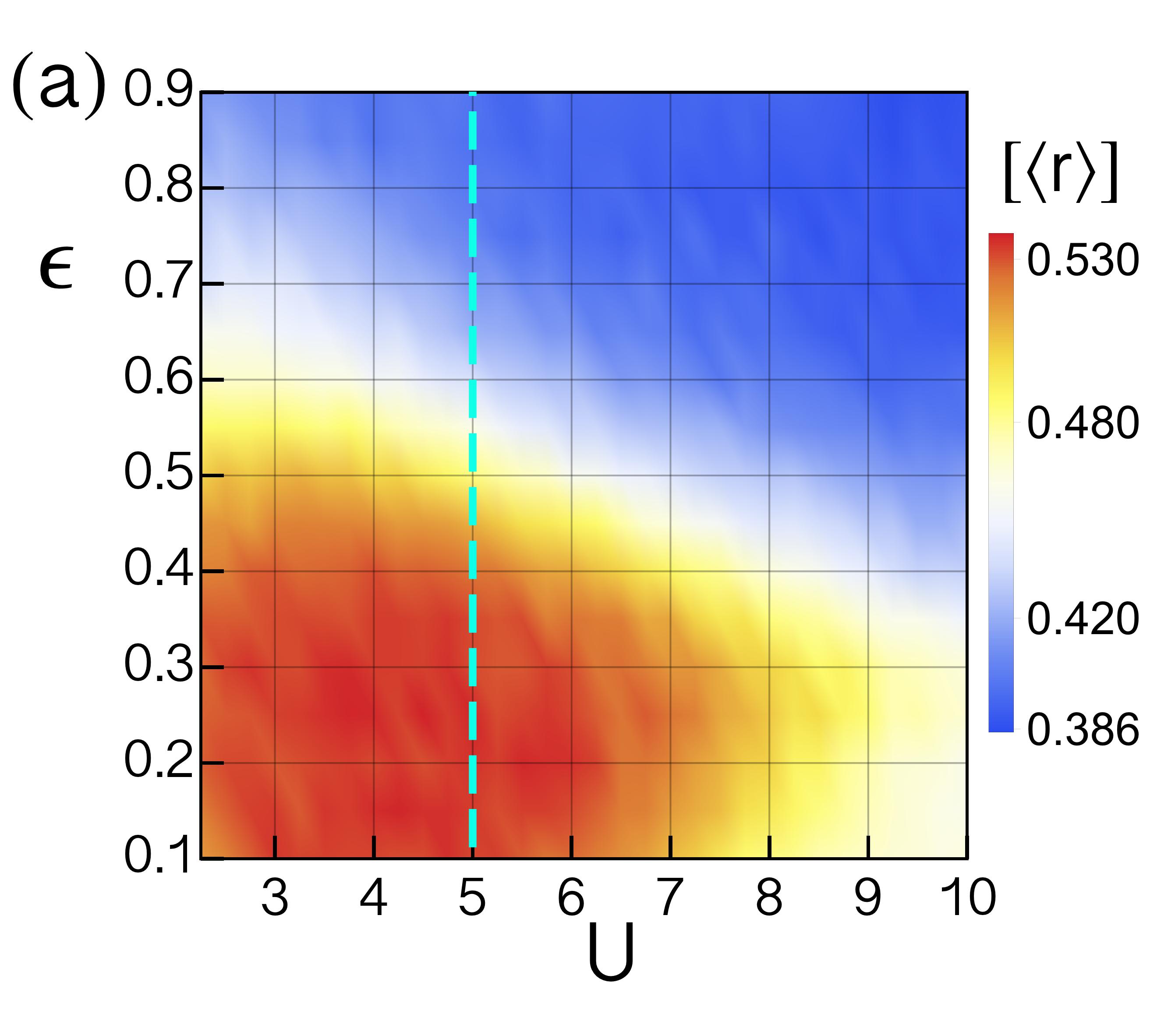}
\includegraphics[trim=0.5cm 0cm 0cm 0cm, scale=0.35]{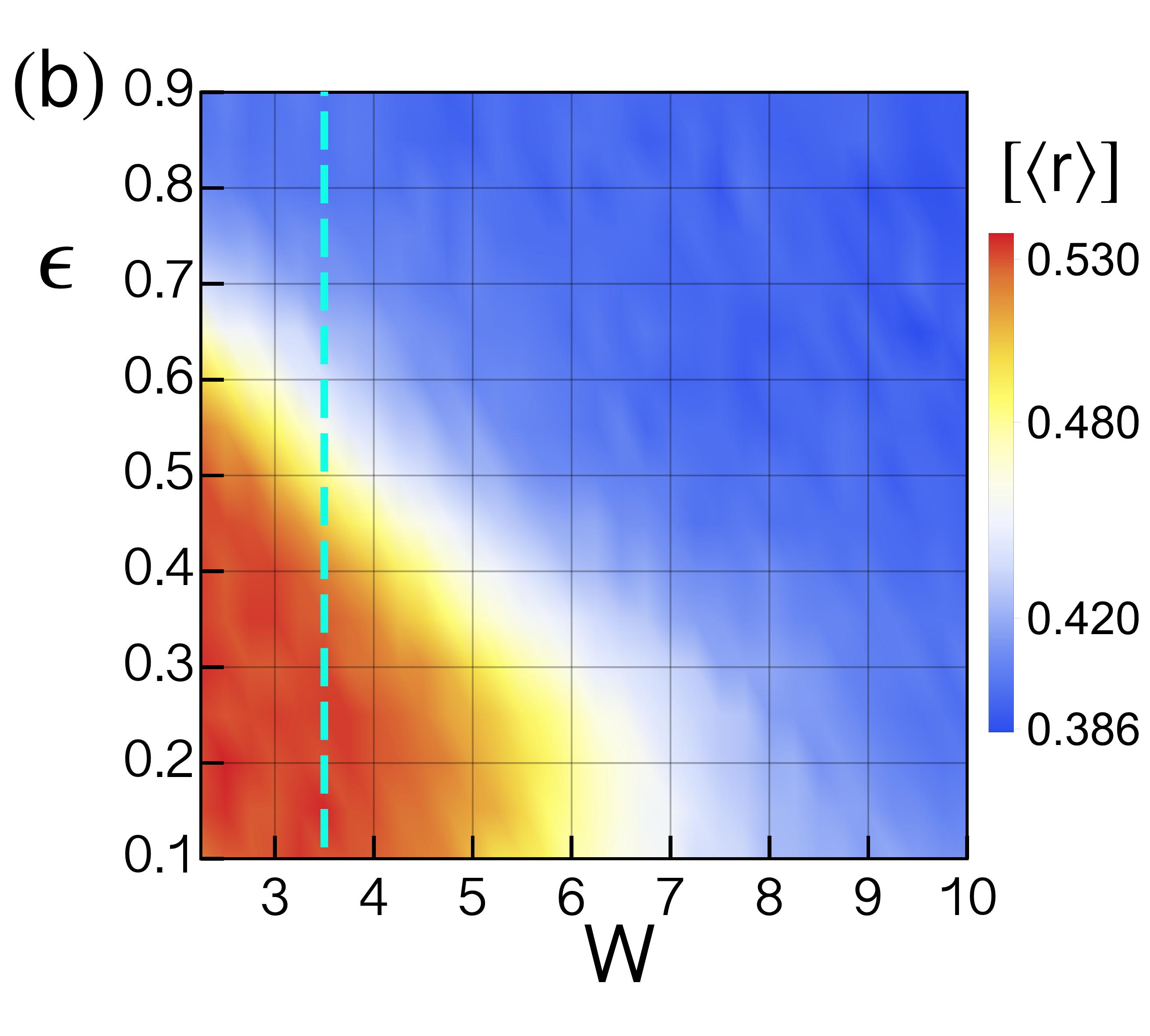}
\caption{Intensity plots of $[\langle r \rangle]$ using correlated disorder as a function of:  (a) interaction strength $U$ with fixed $W=3.5$, and (b) disorder strength $W$ with $U=5.0$. The vertical blue lines denote the particular case for which the other diagnostic calculations are performed and shown in Fig.~\ref{Fig_scal_uncorr}. We used $500$ disorder realizations and $L=14$ for both cases. As with the uncorrelated case, we caution that the mobility edge is not necessarily the white colored region, but rather we estimate it to be within the region for which $[\langle r \rangle]\in (0.42,0.48)$.}\label{Fig_PD_corr}
\end{center}
\end{figure}

\section{Tuning the mobility with correlated disorder}
\label{sec:corrdisorder}

The results and discussion in the previous sections focused purely on uncorrelated disorder. We will now illustrate how, by using correlated disorder, the mobility edge can be made sharper, and can form at lower interaction strengths. The main idea is to implement disorder that selectively couples more strongly to one of  the two dynamical regimes. In our case, to enhance the mobility edge, we want the disorder to couple more strongly to the ferromagnetic states so that the spectral asymmetry that is intrinsic in the system is further increased. The fact that this approach is effective, as shown below, gives further support to the intuitive picture we have presented here concerning the asymmetric dynamics of the XXZ model. Conversely, we could design the disorder to couple more strongly to the anti-ferromagnetic states to make the dynamics more symmetric and thus introduce an opposite trend for the mobility edge. We performed cursory tests that indicate this line of reasoning also works, and we comment on this below, however we will leave a systematic investigation to future work.

\begin{figure}
\begin{center}
\includegraphics[trim=0.cm 0.5cm 0.5cm 0cm, scale=0.358]{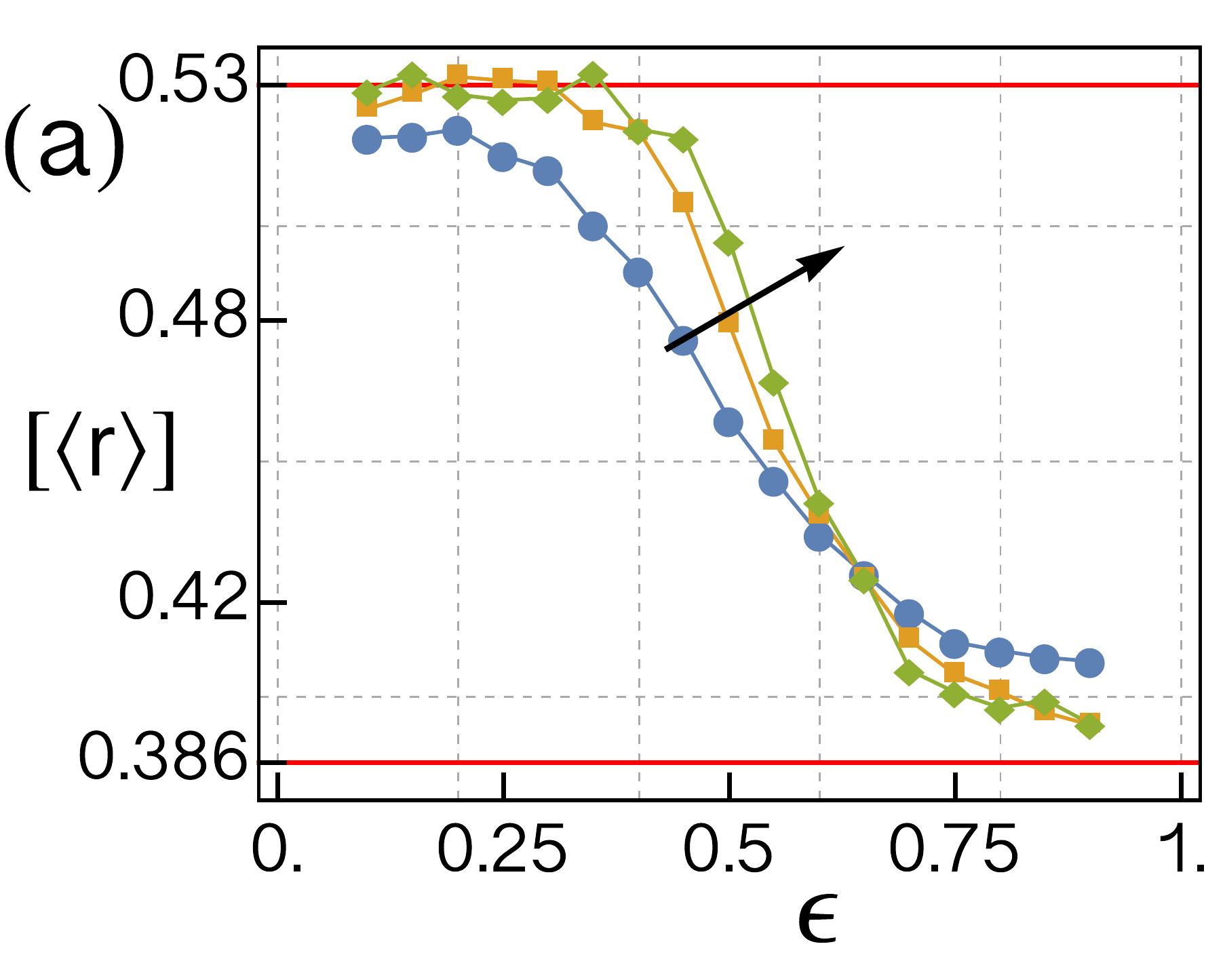}
\includegraphics[trim=0.cm 0.5cm 1.5cm 0cm, scale=0.355]{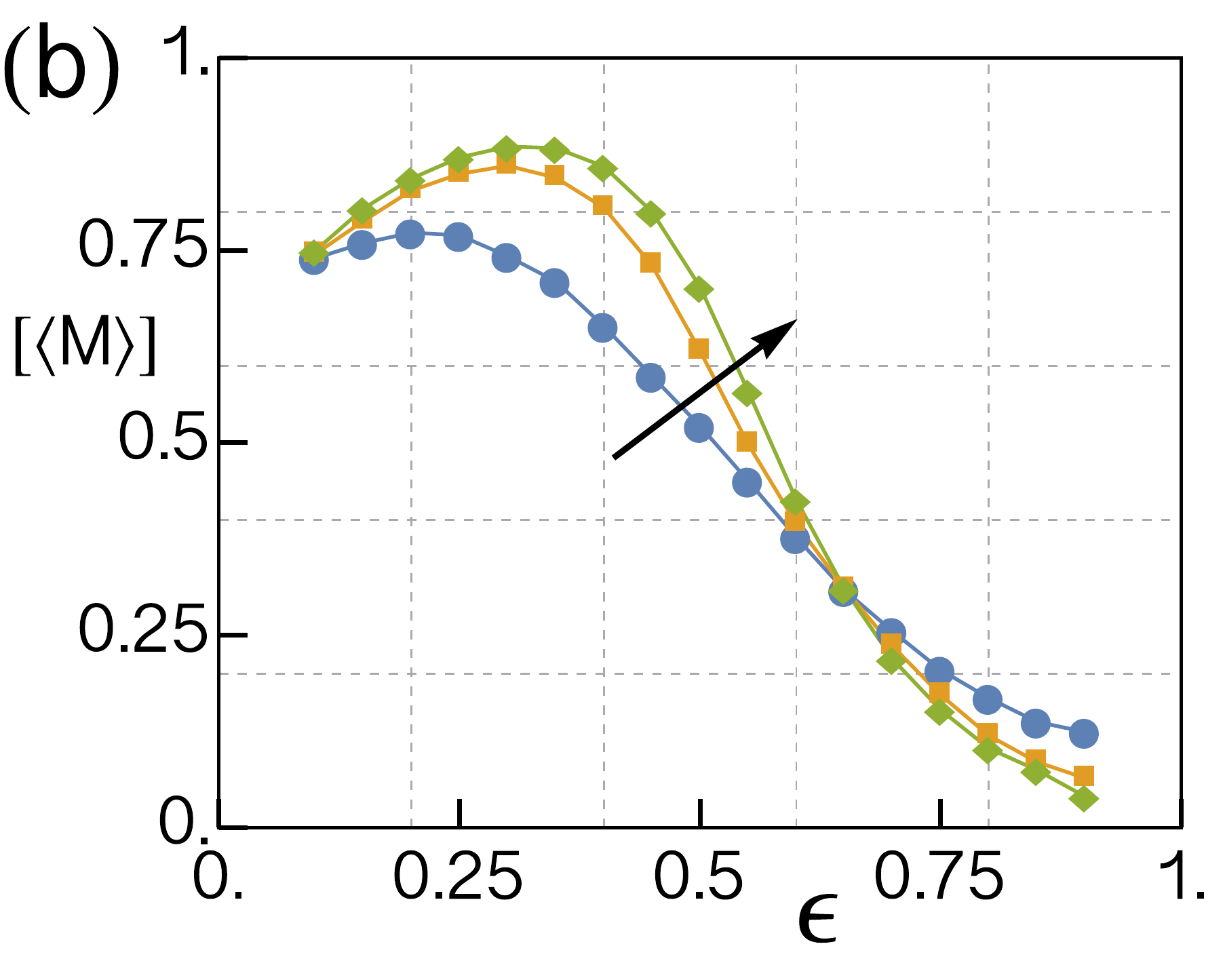}
\includegraphics[trim=0cm 1.5cm 0.cm 0cm, scale=0.355]{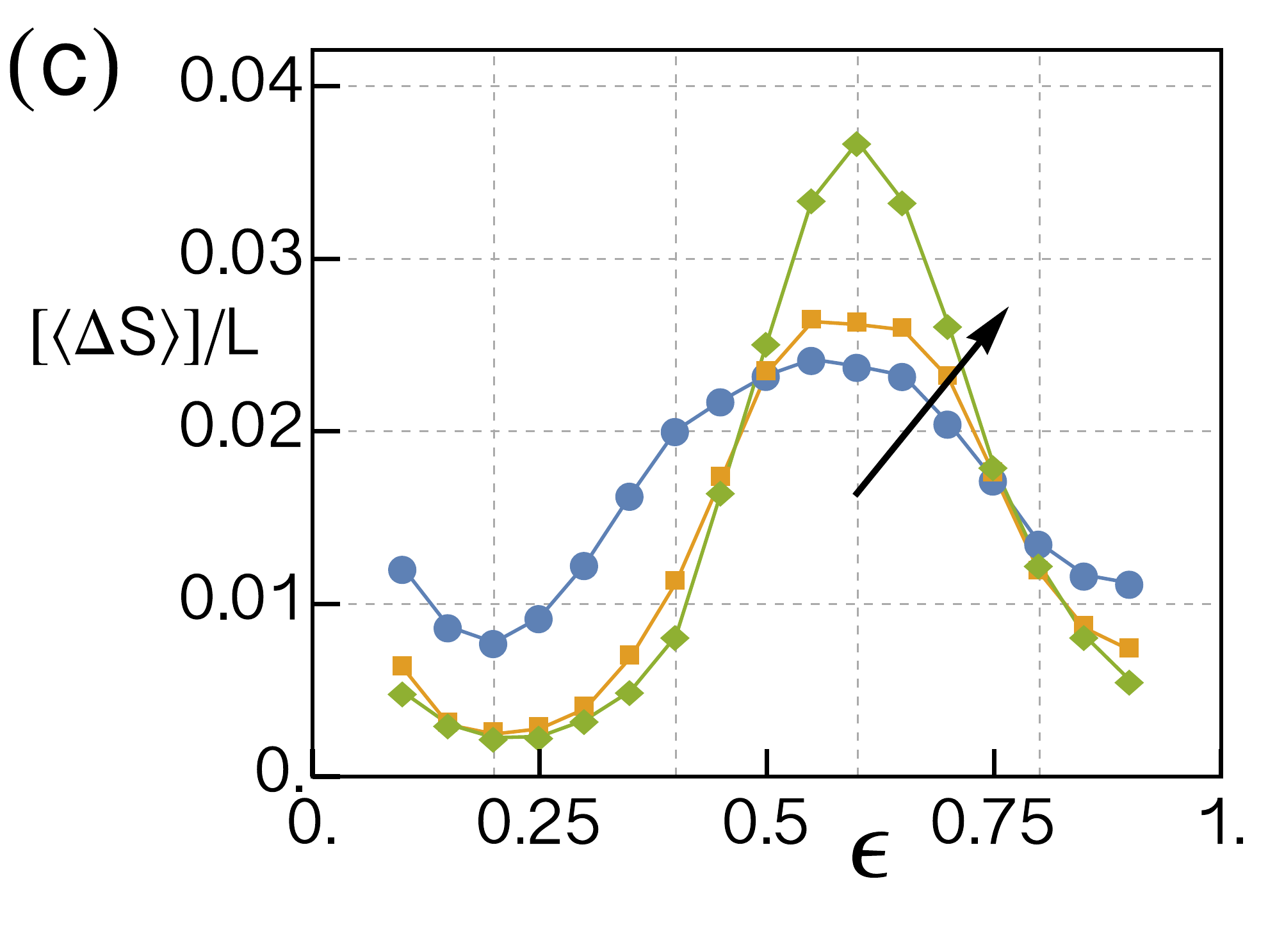}
\caption{Numerical evidence of the mobility edge in the spin-$1/2$ XXZ model using correlated disorder: (a) Statistics of many-body energy level spacings $[\langle r\rangle]$. (b) Fraction of initial spin density modulation which is dynamic $[\langle M\rangle]$. (c) Variance of the entanglement entropy $[\langle \Delta S\rangle]$. Each of the curves in the figures corresponds to the system sizes $L=12\,(\text{blue}),14\,(\text{yellow}),16\,(\text{green})$.  The arrows denote the increase in system size. The parameters used are $(U,W,S_z)=(5.0,3.5,0)$. The number of disorder realizations used was 500 for all the curves, except for the entanglement entropy of the $L=16$ system, for which the number of disorder realizations used was $150$.}\label{Fig_corr}
\end{center}
\end{figure}

To achieve the appropriate type of disorder, we add up sinusoidal potentials of different wavelengths and random shifts:
\begin{equation}
w_n=\frac{2}{\sqrt{L}}\sum^{\frac{L}{2}-1}_{m=1} \tilde{w}_{m} \cos \left(\frac{2\pi m n }{L} +\phi_m\right).\label{wns}
\end{equation}
Here, $\phi_m$ is a random variable that is uniformly distributed in the range $\left[-\pi, \pi\right]$, and the coefficients $\tilde{w}_m$ are real parameters that we choose at will. In particular, we define
\begin{equation}
\tilde{w}_m=
\left\{
	\begin{array}{ll}
		1  & \mbox{if } 0<\frac{2\pi m}{L}<Q, \\
		0 & \mbox{if } Q<\frac{2\pi m}{L}<\pi,
	\end{array}
\right.\label{corrdis}
\end{equation}
where $Q$  is a real number that we tune to improve the signatures of the mobility edge.  For our analysis, we will choose a disorder where we only keep the long wavelength components in Eq.~\ref{wns}, which one would naively expect should couple more strongly to states with ferromagnetic domains. Note that since the Fourier components $\tilde{w}_m$ are normalized to $1$, the overall disorder strength $W$ used in this section cannot be directly compared with the disorder strength used in previous sections. 

The type of disorder we introduced in Sec.\ref{sec:model} has correlations given by $[ w_n w_{n'} ] =\delta_{n,n'}$, which means that it is spatially uncorrelated. In the present case, we obtain
\begin{equation}
[ w_n w_{n'} ]=\frac{1}{L}\left(\frac{\sin\left(\frac{ \pi\left(2M-1\right)(n-n')}{L}\right)}{ \sin \left(\frac{\pi (n-n')}{L}\right)}-1\right),
\end{equation}
where $M$ is the highest integer for which $\tilde{w}_m=1$. This expression explicitly shows that the disorder we are now implementing is spatially correlated. For single-particle Anderson localization it was shown in [\onlinecite{MS2013}, \onlinecite{MS2014}] that this type of correlated disorder generates states that have localization lengths of the order of the system size, effectively leading to regions of delocalized states in the spectrum. This type of disorder is also similar to the correlated disorder that appears in the 1D random dimer model which is known to host extended single-particle states\cite{Dunlap1990}. Additionally, quasiperiodic systems such as the Aubry-Andre model also have suppressed Fourier components which has also been shown to lead to extended states in its energy spectrum \cite{AA1980}. For the many-body case, it turns out this type of disorder also alters the mobility edge, although, we emphasize, due to different reasons as we will discuss below.

We now choose the particular value $Q=3\pi/4$ for the present numerical calculations. With this type of disorder, we computed the phase diagrams in Fig.~\ref{Fig_PD_corr}. The MBL and thermal regions are now much better defined, and the area of the MBL phase has become enlarged. This clearly depicts  that the ferromagnetic states respond more strongly to the model of correlated disorder that we chose, essentially because we have long wavelength disorder which couples strongly to the ``more uniform" ferromagnetic states.

We applied the same mobility edge diagnostics and system size flow as before to this case. We show the results in Fig.~\ref{Fig_corr} for the parameters $(U,W,S^z) =(5.0,3.5,0).$ This case is marked in Figs.~\ref{Fig_PD_corr}a,b as vertical blue lines. As we found for the uncorrelated disorder case, the transition in this case also becomes sharper as the system size increases. Interestingly, the features of each diagnostic have become smoother and clearer compared to the uncorrelated disorder case.

To understand why the use of correlated disorder helps in enhancing the spectral asymmetry of the system in more detail, note that, for a given spin configuration $\{s^z_i\}$, the average Zeeman energy contribution due to the presence of the random field is $\Delta E = W \sum_{i} w_i s^z_i$. The size of this contribution depends on the relationship between the length scales of the variation of spins, and the disorder potential. The energy shift due to the disorder is large when the two length scales are comparable. Heuristically, this gives rise to large shifts in energy which suppresses the resonances responsible for delocalization. In particular, the long wavelength disorder influences ferromagnetic clusters in the negative temperature part of the energy spectrum more significantly. In other words, this type of long wavelength disorder makes states with ferromagnetic correlations more susceptible to MBL by impeding the motion of the excitations in this regime.

We can also consider the opposite case, namely a choice of sinusoidal components of the disorder so that it has a rapidly varying spatial profile instead. In this case,  the antiferromagnetic configurations will couple more strongly to the disorder than the ferromagnetic clusters, and the low energy density part of the energy spectrum will show a greater propensity for localization. However, since there is still a strong tendency for the eigenstates in the ferromagnetic part of the energy spectrum to localize, based on the arguments in Section \ref{sub:quantum_dynamics}, the system still tends to thermalize at an intermediate energy. Unfortunately, due to these two opposing effects, a many-body mobility edge on the antiferromagnetic side of the spectrum is more difficult to observe for the system sizes that we can access, although we did find some indications consistent with this behavior.

\textit{Conclusions.} We have proposed a mechanism that leads to the formation of a many-body mobility edge in a one dimensional model with both uncorrelated or correlated disorder. The mechanism is based on the interplay between strong interactions and the presence of a conservation law that constrains the dynamics of magnon and domain-wall excitations in the ferromagnetic and anti-ferromagnetic regimes of the spectrum, respectively. This leads to an energy-density dependent many-body localization transition which behaves in an asymmetric fashion in the two regimes, reflecting the many-body nature of the transition. We have numerically characterized this mobility edge and furthermore, argued that correlated disorder can be a useful control parameter to enhance the effects of interaction, facilitating the observation of the mobility edge using numerical techniques. Influencing the many-body mobility edge using correlated disorder can also serve as a helpful experimental knob in cold atomic systems where the correlation length of the disorder is controllable. 

\textit{Acknowledgements.} IMS and TLH are supported by ONR award N0014-12-1-0935, the Sloan Foundation, and thank the UIUC ICMT for support. TLH and CRL would like to thank the Aspen Center for Physics and NSF grant $\#$1066293 for hospitality during the conception of this project.

%

\end{document}